# Shape-controlled growth of two-dimensional kagome-lattice colloidal crystals through nanoparticle capping

Rui Huang, Jordan Austin-Frank Wilson, Allen Sun, Artemis Harlow, Zhiwei Li*

Email: zlinano1@umd.edu

Department of Chemistry and Biochemistry, University of Maryland, College Park, Maryland 20742, USA

**Abstract**: Organic capping ligands can selectively bind to crystal facets to modulate growth kinetics and are important in chemical synthesis of inorganic nanocrystals. Using the capping ligands for shape-controlled growth of colloidal crystals is challenging due to the size mismatch of molecules and nanoparticle building blocks. In existing synthetic pathways, colloidal crystal shapes are determined by their thermodynamically favored phases yet controlling their shapes independent of lattice symmetry is vital to study many solid-state properties. Here, we develop a nanoparticle capping strategy to control colloidal crystal shapes and structural heterogeneity. Au bipyramids were used as building blocks and assembled into rhombohedral colloidal crystals driven by DNA hybridization. In (111) planes of the crystals, bipyramids assembled into kagome lattices, featuring structure cavities organized in a hexagonal lattice. The rhombohedral crystals have truncated tetrahedral crystal habits, and the degree of truncation defines the exposed facets and crystal shapes. Our surface capping strategy is to introduce Au–DNA conjugate nanospheres as effective capping agents, which selectively register on the surface vacancies of the kagome facets and resemble the role of organic ligands in classic nanocrystal growth. Such selective capping is driven by maximizing DNA hybridization and leads to slower growth of the (111) kagome facets, changing the crystal shape from three-dimensional truncated tetrahedra to two-dimensional layered microplates with structural heterogeneity and shape anisotropy. This study underpins the importance of capping agents in colloidal crystal growth and inspires effective ways to control the growth kinetics and heterostructures of colloidal crystals.

Controlled chemical synthesis of nanocrystals is fundamental to nanoscience because many physical properties are dependent on nanoparticle shapes and sizes.[1-4] Tuning the physical properties of these solid-state or liquid-phase nanomaterials is central to studying light–matter and matter–matter interactions at nanoscale.[5, 6] Plasmonic nanoparticles, for example, have size-dependent plasmonic resonance frequencies and absorption peak positions which have been broadly tuned across the visible to near infrared spectrum for optical materials engineering and metamaterial designs.[5-8] In catalytic chemical reactions, exposed facets—a group of crystal planes exposed on nanocrystal surfaces—have demonstrated different performances, leading to facet-dependent reaction activity, yield, and product selectivity.[9] Some advanced analytical methods allow probing catalytical reactions occurring on different facets of single nanocrystals, further clarifying the facet dependence of chemical reactions.[10, 11] Many theoretical and experimental analyses have elucidated that the binding affinity of exposed facets with reactants and reaction intermediates is responsible for the facet-dependent activity of nanocrystals.[9, 12] Recent studies on chiral nanocrystals also demonstrate shape-dependent immune responses in biological systems, implying the complexity and importance of nanoparticle shapes in light–matter and matter–matter interactions.[13]

In chemical syntheses, organic ligands that favor selective capping on specific facets of nanocrystals are widely used in shape-controlled nanocrystal growth.[4, 14, 15] The binding affinity between the organic ligands and the favored facets is lower than that of other facets depending on atom types and arrangement in crystal planes.[16-18] Such selective capping stabilizes the facets and slows down deposition of atoms that leads to facet growth. As a result, fast-grown facets will disappear while ligand-capped facets remain to form organic–inorganic interfaces.[19] For example, polyvinylpyrrolidone (PVP) and trisodium citrate (TSC) prefer to bind to (100) and (111) facets of Ag, respectively, leading to cubic and octahedral nanocrystals that are terminated by the capped facets.[20, 21] If twinned seeds were used, introducing PVP and TSC as capping ligands produces Ag nanoplates and nanowires terminated by the (100) and (111) facets, respectively.[22, 23] Choosing organic ligands also prevent the nanocrystals from physical aggregation and chemical transformation. Therefore, capping ligands not only regulate the growth kinetics of nanocrystals, but also determine surface properties.[24]

Compared with nanocrystal growth, controlling the shapes and habits of colloidal crystals—crystals made of nanoparticle building blocks—is challenging. Introducing the organic ligands in nanocrystal synthesis for colloidal crystal growth is difficult due to the size mismatch between the ligands and nanoparticle building blocks, in which the ligands' binding affinity with distinct colloidal crystal facets is trivial at the nanoscale. Although many methods have been developed to synthesize colloidal crystals with distinct crystal structures and symmetries, the crystal shapes are usually coupled to the thermodynamic crystal phases.[25, 26] As a result, existing colloidal crystal synthesis produces either arbitrary shapes due to the lack of thermodynamic constrains[27] or shapes dictated by the thermodynamically favored crystal phases (e.g., cubes for single cubic lattices,[25] rhombic dodecahedra for body-centered cubic lattices,[28] and octahedra for face-centered cubic lattices[25]).[29, 30]

Here, we introduce selective capping of nanoparticles on colloidal crystal facets in DNA-mediated co-assembly of nanospheres and bipyramids. Consider non-covalent bonds between atoms and organic surfactants, the binding affinity is dictated by atom arrangement in crystal facets, which is responsive to selective capping. Several studies on DNA-mediated assembly have come to appreciate that DNA binding between nanoparticle building blocks is similar in many ways to those employed in atom bonds, including valences, bond directionality and symmetry.[31-34] Imagining nanoparticles conjugated with DNA strands, we elucidate that the binding affinity of nanoparticles of colloidal crystals is dependent on facets that are made of nanoparticle building blocks in distinct positional and orientational orders. Such selective binding of nanoparticles to one facet over other facets resembles the capping ligands binding to nanocrystal facets, thus regulating the growth kinetics of colloidal crystals and dictating the shapes and habits independent of crystals' thermodynamically stable phases.

Au bipyramids of 115 nm were modified with anchor DNA strands and used as building blocks to synthesize the colloidal crystals (**Fig. s1**). Linker DNA strands were then added to pair with the anchor strands while, on another linker end, GCGC self-complementary sequences will hybridize to drive the nanoparticle assembly (**Fig. s2**). This assembly occurs in a slow cooling process (0.1 $^{o}$C/20 minutes). Based on SEM images, we observed the formation of polyhedral crystals with exposed flat facets and well-defined sharp edges. **Fig. 1a** shows a crystal with low degree of truncation along four tips of a tetrahedron, leading to eight exposed facets. As the truncation degree increases, the new facet becomes larger (**Fig. 1b**) while the initial four facets are decreasing. At critical truncation in **Fig. 1c**, the crystal adopts an octahedral shape with eight equally large facets. In each facet, we observed different orders and symmetries, including kagome facets in **Fig. 1d** and lateral facets in **Fig. 1e**. The truncation-induced formation mechanism of different shapes is demonstrated in **Fig. 1f**, which explains the crystal shapes observed in **Figs. 1a-1c**. The initial tetrahedron has three exposed (100) side facets and one (111) bottom facet. The truncation occurs along two crystalline directions <100> and <111>, leading to the formation of three new (100) facets and one new (111) facet. At the critical truncation, we observed the formation of octahedron with six (100) and two (111) facets, corresponding to the crystal observed in **Fig. 1c**.

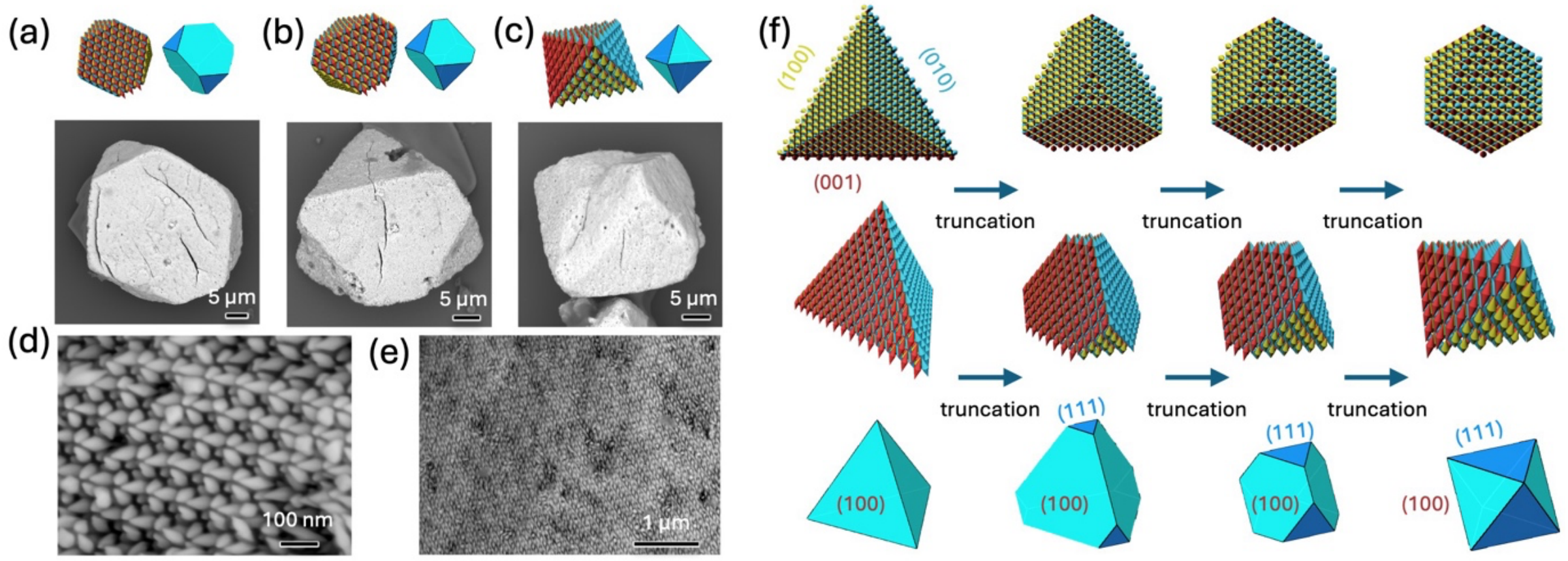

**Figure 1. (a-c)** SEM images of tetrahedral colloidal crystals with different degrees of truncation. SEM images of **(d)** (111) facets and **(e)** (100) facets. **(f)** Schematic illustration of truncation of the tetrahedral crystals.

For the Au bipyramid assembly, three-dimensional polyhedral crystals were observed in SEM images (**Fig. 2a**). To verify the positional and orientational order of the Au bipyramids, we prepared 50 nm-thick slides of one truncated tetrahedral crystal (**Fig. 2b**) and examined the local orders of Au bipyramids using transmission electron microscopy (TEM). In **Fig. 2c**, the dark field

TEM image shows the cross section of Au bipyramids assembled in the kagome lattices. As shown in the scheme in **Fig. 2c**, three Au bipyramids assembled into a triangular subunit through tip-tip interactions driven by DNA hybridization. The subunit further packs into a hexagonal 2D lattice in one facet, leading to the formation of kagome lattices. The polyhedral colloidal crystals are made of multilayer stacking of such kagome lattices. Due to the elongated pyramidal structures, three Au bipyramids tilt such that their surface is parallel to maximize hybridization of DNA between adjacent two bipyramids. The regular pentagonal cross section of the Au bipyramids leads to an 108° internal angle, which is responsible for three bipyramids packing into the triangular subunits (**Fig. s3a**). While the subunits contain three bipyramids with top parallel bonding, the connectivity between adjacent subunits is occurring through bottom parallel bonding, leading to the formation of kagome lattices as illustrated in **Fig. s3b**. In kagome planes, the lattice has three-fold rotational symmetry and regularly distributed vacancies that serve as the bonding sites for the lateral growth of another kagome plane (**Fig. s3c**). Therefore, the accessibility of the active vacancy sites is critical for continuous colloidal crystal growth.

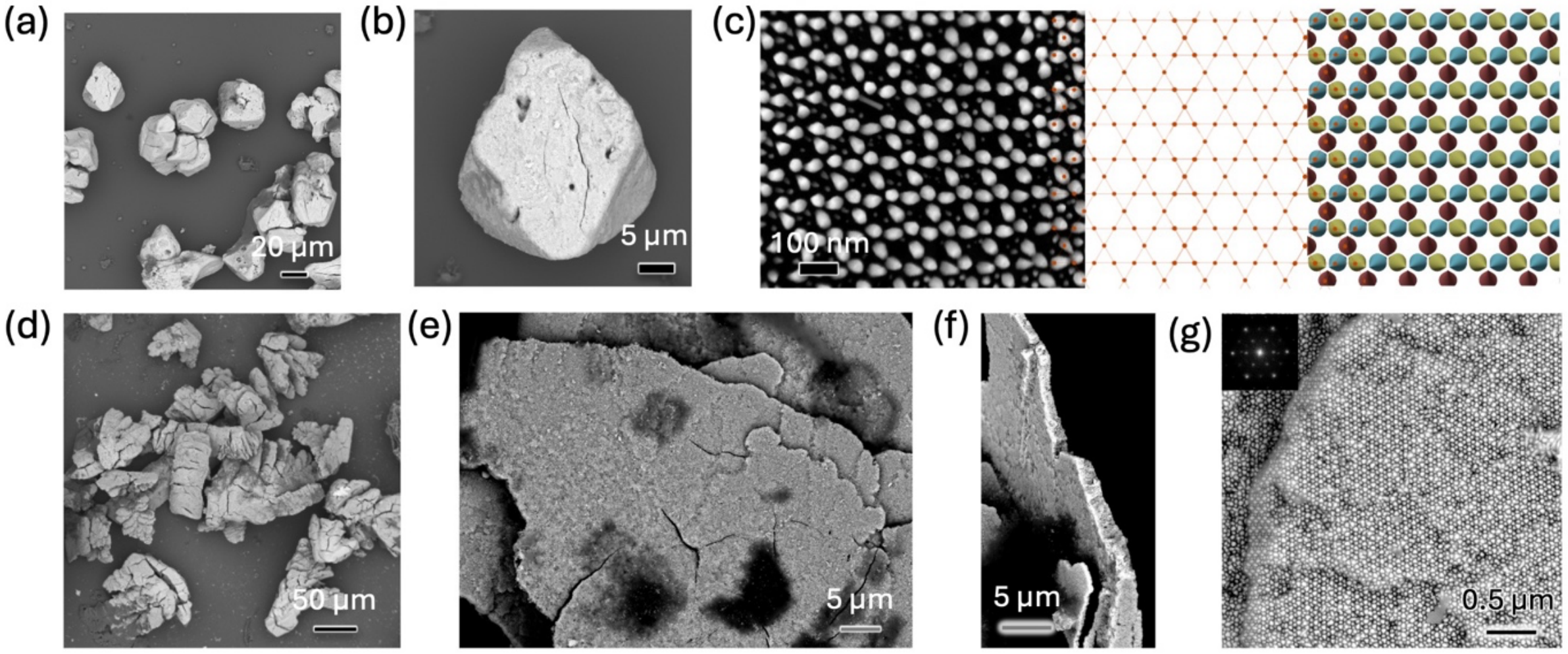

**Figure 2. (a)** SEM image of polyhedral colloidal crystals assembled from bipyramid nanoparticles. **(b)** SEM image of single-crystal truncated tetrahedron. **(c)** Cross-sectional TEM image of in-layer packing of bipyramids into kagome lattices. The positional and orientational orders of the bipyramids is illustrated in the middle and right lattices, respectively. **(d)** SEM image of 2D layered colloidal crystals by co-assembling bipyramids with nanospheres. **(e)** Low-magnification top, **(f)** lateral and **(g)** high-magnification top SEM image of one single-crystal, layered structure. Inset in (g) is the FFT of the SEM image.

Based on the unique structures of the crystals, it is possible to regulate the growth of the kagome facets by controlling the accessibility of the active vacancy sites. In nanocrystal synthesis, the binding energy difference is the driving force for the selective capping of molecule ligands on crystal facets. It is therefore important to search for nanoparticles having different binding energy with colloidal crystal facets. To this end, we introduced spherical Au nanoparticles with an average diameter of 80 nm as capping agents considering their shape and size matching with the vacancies and ease of preparation. The Au nanospheres were modified with the same DNA anchor strands and co-assembled with the bipyramids. We observed an interesting crystal habit change from 3D polyhedral shapes to 2D layered structures (**Fig. 2d**). Different from the well-defined facets in the polyhedral crystals, the layered crystals have large, flat surfaces (~ 50 μm in length, **Fig. 2e**) and are very thin (~ 1 μm in thickness, **Figs. 2f, s4a,** and **s4b**). We observed complex structures with six-fold rotational symmetry on the surface of the colloidal crystal sheet, demonstrating the good crystallinity and structural integrity (**Fig. 2g**). Such significant shape changes are induced by the addition of Au nanospheres in the DNA-mediated assembly, making the two-dimensional layered shapes more thermodynamically preferred in the same assembly condition.

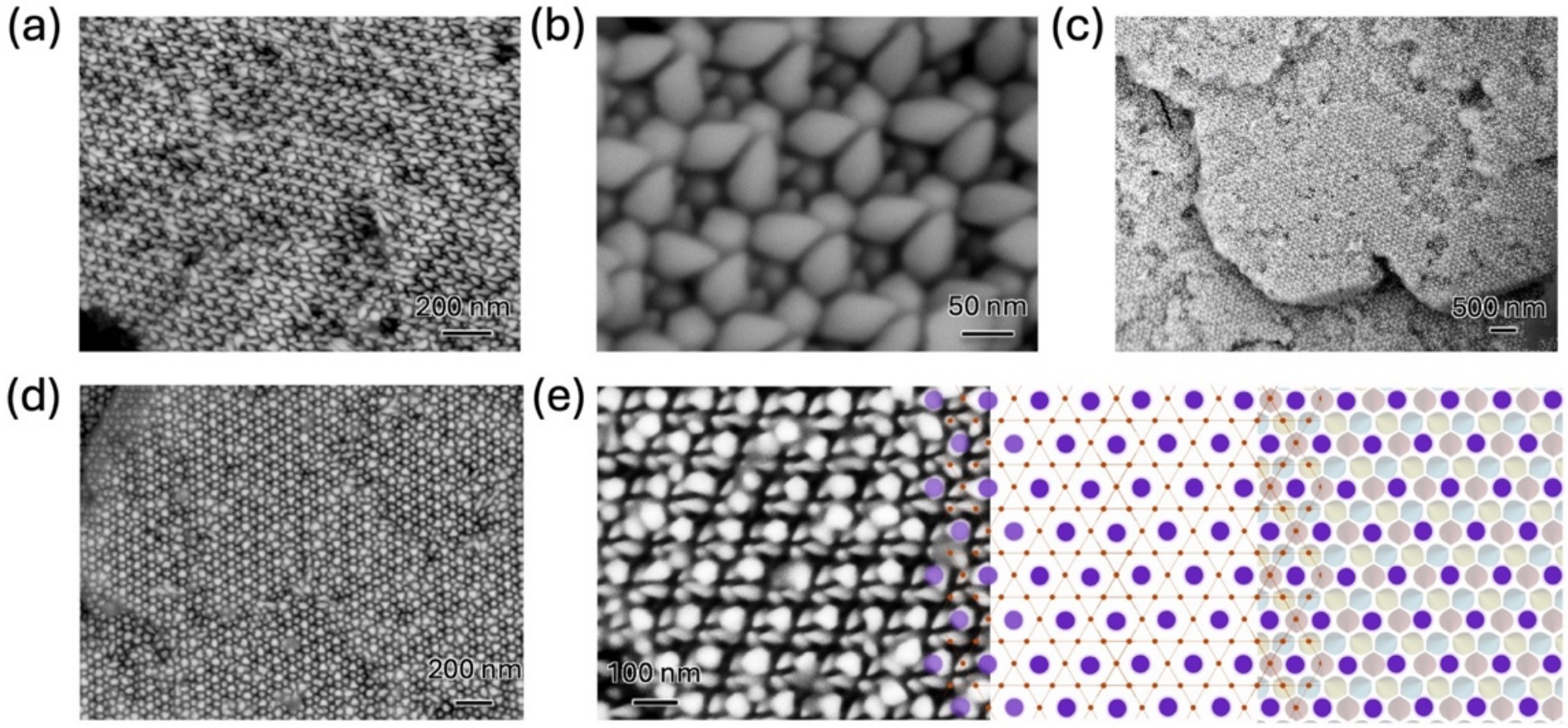

**Figure 3. (a-b)** SEM images of the (111) facets of polyhedral colloidal crystals. Au bipyramids pack into kagome lattices without capping of nanospheres. **(c-d)** SEM images of the (111) facets of colloidal crystal sheet. The large vacancies of kagome lattices are capped by nanospheres. **(e)** High-magnification SEM image of the (111) facets. The positional order of the bipyramids and nanospheres is illustrated on the right scheme.

To investigate whether the Au nanospheres serve as capping reagents, we carefully examined the SEM images and elucidated positional orders of nanoparticles. Assembling Au bipyramids produces polyhedral colloidal crystals made of stacked kagome lattices. The exposed (111) facets show the packing of bipyramids into kagome lattices (**Fig. 3a**), which features the subunits made of three bipyramids and their propagation into hexagonal superlattices with periodic surface vacancies (**Fig. 3b**). The SEM image in **Fig. 3c** further reveals how the structure vacancies form in the (111) kagome facets. That is, six bipyramids from three neighboring subunits form large vacancies through parallel, top tip–tip bonding. And three bipyramids from another three neighboring subunits form small vacancies through parallel, bottom tip–tip pairing. The two sets of vacancies have the same hexagonal order within one kagome lattice plane. The SEM images in **Figs. 3d**, **3e, s4c,** and **s4d** demonstrate that the vacancies disappeared and were instead filled with Au nanospheres. Comparing the SEM images in **Fig. 3c** and **3f**, we observed that the 80 nm Au nanospheres assembled into the large surface vacancies that are formed by six bipyramids. In **Fig. 3c**, the bipyramidal tips on the second layer can be observed through the large surface vacancies while in **Fig. 3f**, the Au nanospheres assemble in the local large vacancies into a hexagonal lattice. These observations suggest that the Au nanospheres of 80 nm can selectively assemble and register in the large surface vacancies in the kagome facets, leaving the small vacancies open and unoccupied. During assembly, the large vacancies surrounded by six bipyramids are reactive and responsible for the packing of another kagome lattices into 3D colloidal crystals. Once nanospheres are added, the selective registration of nanospheres on the active vacancies of the kagome facets compete with the assembly of Au bipyramids. The better size and shape match of the nanospheres with the reactive vacancies allows selective, favored binding, which is driven by maximizing DNA hybridization. This sequence of event leads to the preferred capping of nanospheres on the kagome facets through vacancy registration, slows down the normal growth of the kagome planes, and produces 2D layered crystals as thermodynamically favored assemblies.

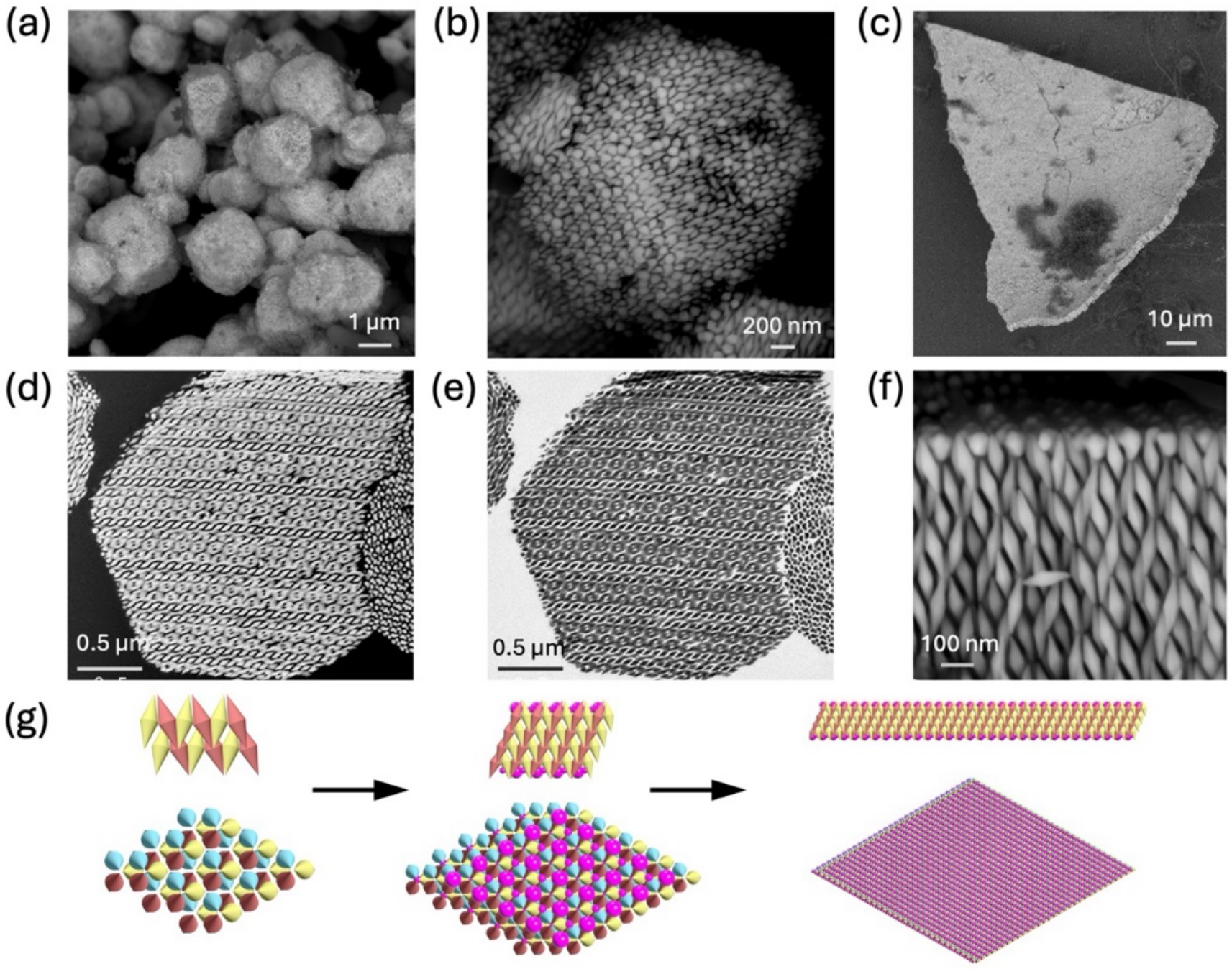

**Figure 4.** SEM images of **(a)** nanocrystals, **(b)** a single nanocrystal in early stage of growth, and **(c)** one colloidal single-crystal sheet after shape-controlled growth. **(d)** Dark-field and **(e)** bright-field cross-sectional TEM images of one colloidal crystal. **(f)** SEM image of the lateral facets of the 2D colloidal crystals. **(g)** Schematic illustration of the selective capping of nanospheres on the (111) facets of the crystals, leading to the formation of the 2D, layered colloidal crystals.

Since Au nanospheres and bipyramids were added at beginning of the co-assembly, understanding the interactions between these two types of building blocks is critical to determine the formation mechanism and assembly pathways. We quenched the assembly at early stage of the slow cooling to investigate the intermediate crystal structures with the presence of Au nanospheres. As shown in **Fig. 4a**, the SEM image shows small colloidal crystals of a few micrometers. Although the reaction was quenched during colloidal crystal growth, we saw high-quality crystals with broad size distribution and orientations. These small crystals are analogy to tiny nuclei in nanocrystal growth if one considers nanoparticles as building blocks. Further analyzing these colloidal "nuclei" reveals that the Au bipyramids have well-defined, distinct positional and orientational orders in exposed facets, which can be observed on nuclei with various orientations. In nanocrystal growth, capping ligands can selectively bind to and stabilize one type of crystal facet, thus slowing down the deposition of reduced atoms. Directly observing the preferred binding is challenging considering the small sizes of organic molecules. Here, this experiment demonstrates selective capping of nanospheres on colloidal crystals and elucidates how the selective capping regulates the growth of 2D layered colloidal crystals. As shown in **Fig. 4b**, we focused on the one single-crystal nucleus from the early quenched sample and found that the top (111) facets feature regularly packed Au nanospheres that assembled in the large vacancies. On the contrary, the side facets are clean and contain Au bipyramids in well-defined positional and orientational orders. Although the size varies, the nuclei are faceted and single crystals. The selective assembly of 80 nm Au nanospheres was demonstrated on the top (111) facets, leaving the side facets uncapped and available to further bipyramid deposition. In the co-assembly of nanospheres and bipyramids, the large vacancies are preferentially occupied by Au nanospheres, making the interlayer bonding sites physically occupied and inaccessible to the bipyramids. In the slow cooling process, the side facets not attached by Au nanospheres are still active and accessible to bipyramid attachment driven by DNA hybridization. Such a difference in force dynamics leads to preferred growth along the lateral direction and limited growth along thickness direction, finally resulting in 2D layered colloidal crystals with uniform thickness (**Fig. 4c**). To verify the selective bonding of Au nanospheres, we further examined the cross section of one nucleus under TEM. As shown in **Fig. 4d** and **4e**, the dark and bright field TEM images confirm the single crystallinity of the nucleus, which exhibits structural uniformity inside the nucleus. We observed the presence of Au nanospheres only on the kagome facets, confirming that the selective capping only occurs on the kagome lattices. Moreover, the SEM image in **Fig. 4f** also confirms that Au nanospheres are well-positioned in the large vacancies on the exposed facets.

The self-complementary DNA strands will hybridize when two Au nanoparticles, either nanospheres or bipyramids, are in contact. During the slow cooling, the DNA-mediated assembly can be predicted by maximizing the DNA hybridization between neighboring building blocks.[29, 35-37] Therefore, thermodynamically stable crystals are expected to have maximum DNA hybridization and therefore the most stable assemblies feature maximum surface–surface overlaps between building blocks. This consideration combined with our observation suggests a reasonable growth pathway to the 2D layered crystals, which is driven by DNA hybridization and regulated by the selective capping of Au nanospheres on the kagome lattices. The formation of the nuclei is initiated by bipyramid–bipyramid interactions due to the large overlap between DNA-grafted plain surfaces (left images in **Fig. 4g**). It is highly possible that the Au bipyramids first assemble into small crystals with a few layers of kagome lattices, during which large surface vacancies form on the surface. Since the vacancy size is comparable to the diameter of the Au nanospheres, they serve as preferred binding sites for Au nanospheres, which is experimentally evidenced by the SEM images in **Fig. 4f**. After Au nanosphere registration, the facet and structures will be stabilized by the DNA interactions thanks to the large surface overlap between individual Au nanospheres and the vacancies. This step initiates the selective capping of Au nanospheres on the top (111) crystal facets, which stabilizes the facets and slows down deposition of bipyramids (middle panels in **Fig. 4g**). The preferred deposition of Au bipyramids along the side facets over the top facets produces 2D layered colloidal crystals (right panels in **Fig. 4g**) with single crystallinity (**Fig. s5**).

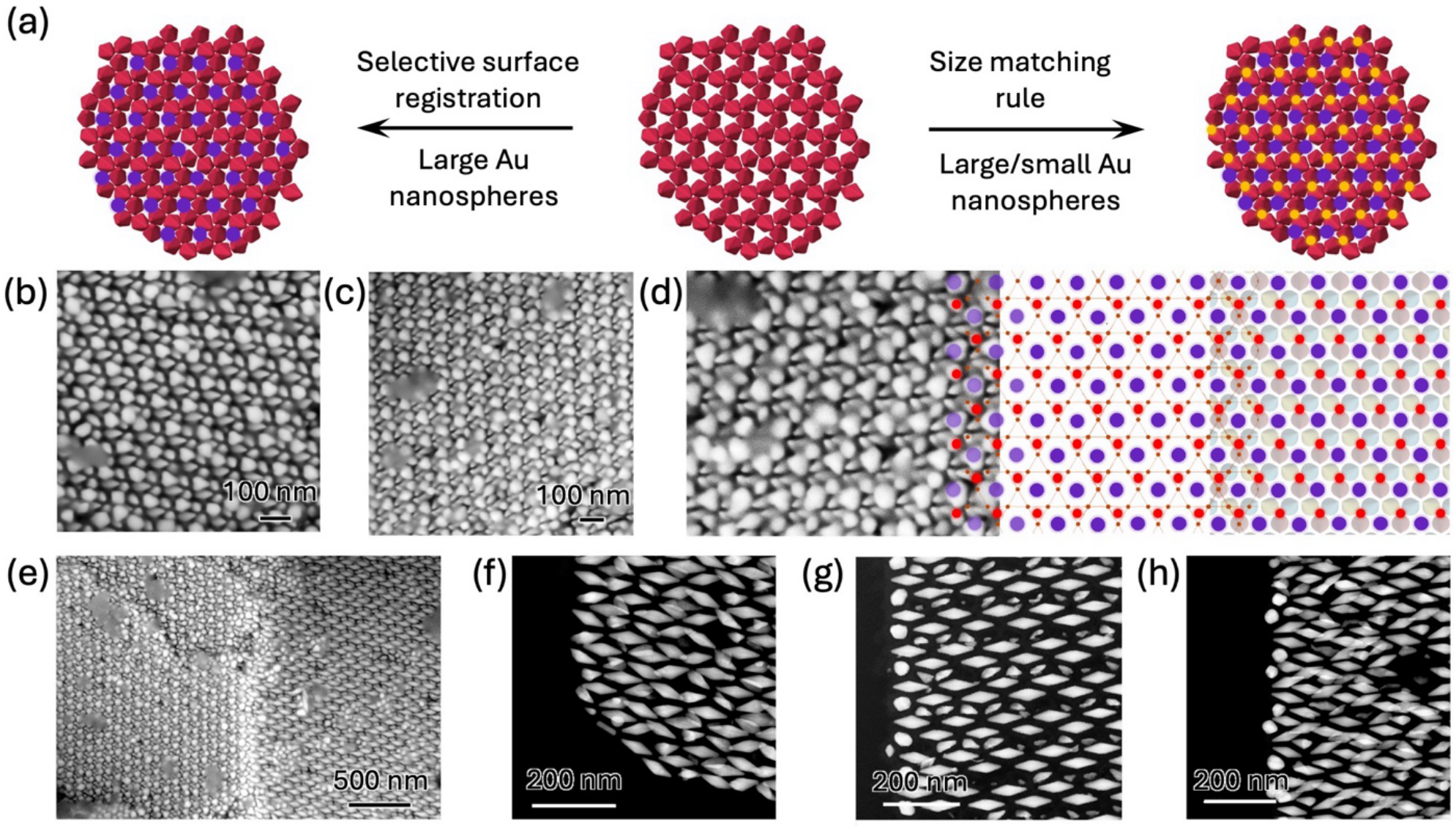

**Figure 5. (a)** Schematic illustration of surface-selective registration of nanospheres of two different sizes on the (111) facets of the 2D colloidal crystals. SEM images of the (111) facets of colloidal crystals by using Au nanospheres of **(b)** 80 nm, **(c)** 80 nm and 20 nm as capping reagents. **(d)** Schematic illustration of the positions and positional order of the 80-nm and 20-nm Au nanospheres within the (111) facets. **(e)** SEM image showing the high selectivity of surface registration of binary nanospheres. Dark-field cross-sectional TEM images of **(f)** non-capped, **(g)** large nanosphere-capped, and **(h)** binary nanosphere-capped colloidal crystals.

Further analysis of the 2D layered crystals suggests that 80 nm Au nanospheres exhibit high selectivity in registering in large vacancies over small vacancies in the top (111) facets. Considering the presence of small and large vacancies and imagining the structures as being driven by maximizing DNA hybridization, it is possible to form complex heterostructures by applying the size matching rule. So, we introduced Au nanospheres of 20 nm and 80 nm as secondary building blocks (**Fig. s6**) and studied co-assembly with bipyramids using the same slow-cooling method. Based on the size matching rule, we expected size-dependent surface registration of Au nanospheres (**Fig. 5a**). If only large Au nanospheres are used, we observed their exclusive registration on large vacancies (**Fig. 5b**), while all small vacancies remained open. By controlling small Au nanospheres concentration, we realized highly selective registration of the 20 nm Au nanospheres into the small vacancies (**Figs. 5c**). Specifically in **Figs. 5d** and **s7**, we observed hexagonal packing of 20 nm and 80 nm Au nanospheres into hexagonal lattices, which are separately registered in the small and large vacancies, respectively. The size-dependent assembly only occurs in the kagome planes, all other exposed facets are clean. We could see some nanospheres randomly attached on the side facets without order while the whole facet remains accessible (**Fig. 5e**). To further examine the crystallinity and registration of nanospheres, thin layers of the colloidal crystals were prepared to image the cross sections under TEM. As demonstrated in **Figs. 5f**, **5g**, and **5h**, the packing of bipyramids inside the crystals are in high order, with limited point defects. TEM image in **Fig. 5f** demonstrates clear and open facets when bipyramids were the only building blocks. When 80 nm Au nanospheres are used as capping agents, the large vacancies were occupied, leading to periodic arrangement of the nanospheres (**Fig. 5g**). In **Fig. 5h**, we observed two sets of nanospheres located in upper and lower positions in the kagome facets, which correspond to the small and large vacancies on the (111) facets in co-assembling Au nanospheres with bipyramids. These TEM images clarify the size-dependent, highly selective surface registration of Au nanospheres on kagome lattices.

In summary, we introduce the selective capping effect of nanoparticles for shape-controlled growth of colloidal crystals with high structural complexity and hierarchies driven by DNA hybridization. DNA-modified Au bipyramids have non-space-filling shapes and assemble into kagome lattices, leading to the formation of 3D polyhedral colloidal crystals with well-defined facets as well as crystal habits. Adding Au nanospheres that are modified with the same DNA strands changes the crystal habits from 3D truncated tetrahedron to 2D layered microplates. Such shape changes are induced by exclusive registration of the Au nanospheres on the vacancies of the kagome facets, which stabilizes the facet and slows down the bipyramid deposition and normal crystal growth on the facets. This selective surface capping on colloidal crystal facets and its induced shape control resembles the surface ligand chemistry in nanocrystal growth that occurs at organic-inorganic interfaces of individual nanoparticles. While binding strength depends on molecular structures of capping ligands in nanocrystal growth, this study suggests that selective surface capping of nanospheres is realized based on their geometry matching with the vacancies of the kagome facets. This capping strategy in colloidal crystals paves effective ways to control the growth kinetics and crystal habits of superstructures made of nanoparticle building blocks. It also provides a way to synthesize heterogeneous colloidal crystals with controlled structural heterogeneity.

**Acknowledgements**

This work was supported by the startup funds from the University of Maryland, College Park.

## Supporting Information

## Shape-controlled growth of two-dimensional kagome-lattice colloidal crystals through nanoparticle capping

Rui Huang, Jordan Austin-Frank Wilson, Allen Sun, Artemis Harlow, Zhiwei Li*

Email: zlinano1@umd.edu

Department of Chemistry and Biochemistry, University of Maryland, College Park, Maryland 20742, USA

### Synthesis of Au bipyramids and nanospheres

The synthesis of Au bipyramids is based on a previously reported method. Penta-twinned Au nanoparticles were first synthesized and used as seeds to grow Au bipyramids. At ambient condition, 10 mL of $HAuCl_4$ (0.25 mM), cetyltrimethylammonium chloride (CTAC, 50 mM), and sodium citrate (5 mM) was prepared and mixed in an aqueous solution. 0.25 ml of $NaHB_4$ (25 mM) was freshly prepared and added into the solution under stirring. The color of the solution changed from light yellow to brownish, demonstrating the formation of small Au nanoparticles. The reaction was continuing at the ambient temperature for 30 minutes and then was moved to a thermomixer for aging at 80$^o$C for 90 minutes. During the high-temperature aging, the Au nanoparticles' size increases, and the solution showed a color change from brown to red. The seed solution was stored at room temperature for growing Au bipyramids. In the seeded growth step, the temperature was set at 30 $^o$C and the growth solution was prepared by mixing 100 ml of CTAB (100 mM), 5 ml of $HAuCl_4$ (10 mM), 1 ml of $AgNO_3$ (10 mM), 2 ml of HCl (2 M), and 0.8 ml of AA (0.1 M) in sequence. 200 μL of seed solution was added into the growth solution, and the seeded growth was performed 2 hours at 30 $^o$C. The bipyramids of 115 nm were washed with water twice to remove excessive reactants and dispersed in 10 mL of water for DNA functionalization.

The preparation of Au nanospheres of 20 nm and 80 nm were prepared by a seeded growth method. The Au seeds (~3 nm) were prepared by mixing 1 mL of $HAuCl_4$ (5 mM), 1 mL of TSC (5 mM) with 18 mL of $H_2O$ in a flask. Under magnetic stirring, 0.6 mL of freshly prepared $NaBH_4$ solution (0.1 M) was quickly added. After 4 hours, the solution was used as seed solution. In the seeded growth step, a growth solution was prepared by mixing 500 μL of CTAB (20 mM), 250 μL of l-ascorbic acid (0.1 M), 200 μL of KI (0.2 M), 60 μL of $HAuCl_4$ (0.25 M), and 2 mL of $H_2O$. 1600 μL and 20 μL of seed solutions were added for growing 20-nm and 80-nm Au nanospheres, respectively. After 10 minutes, the Au nanospheres were collected by centrifugation and dispersed in water for DNA modification.

### DNA functionalization on Au bipyramids and nanospheres

The DNA anchor strands are thiolated oligonucleotides. 50 nmol of the anker DNA strands was used to modify each 1-mL of the Au bipyramids stock solution. In the first step, thiolated DNA anchor strands were reduced with dithiothreitol (0.1 M in 170 mM phosphate buffer) for 1 hour at ambient condition and purified by desalting columns. During the reduction, 1 mL of Au bipyramids or nanospheres solutions were centrifugated and supernatant was carefully removed. In the second step, the reduced and desalted DNA anchor strands (50 nmol) were added into the Au nanoparticles, followed by adding 0.1 mL of SDS (0.1%) and PBS (0.1 M). The final volume of the mixture was fixed at 1 mL by adding water to the DNA-particle solution. The solution was sonicated to fully disperse the nanoparticles and placed in a thermomixer overnight. In the third step, NaCl (2M) was added to promote ligand replacement on Au nanoparticle surfaces. 25.6, 27.0, 58.5, 65.4, 73.5, and 83.3 μl of NaCl (2M) was added each 30 min to the DNA-particle solution. The mixture was further shaken in the thermomixer for 24 hours and washed with 0.01% SDS twice.

**Assembly of colloidal crystals by slow cooling**

The DNA-particle conjugates were dispersed in 100 μL of NaCl (0.5 M) solution with SDS (0.01%) and PBS (0.01 M). DNA linker strands (10 nmol) were added into the nanoparticle solution. The assembly was performed in a thermal cycler. The cooling rate for temperature between 70°C and 66°C, 66°C and 45°C, and 45°C and 25°C was 0.1°C/10 min, 0.1°C/20 min, and 0.1°C/10 min, respectively.

**Characterization**

SEM images were acquired using Tescan GAIA/XEIA FEG SEM and TEM images were acquired using JEM 2100 LaB6 TEM.

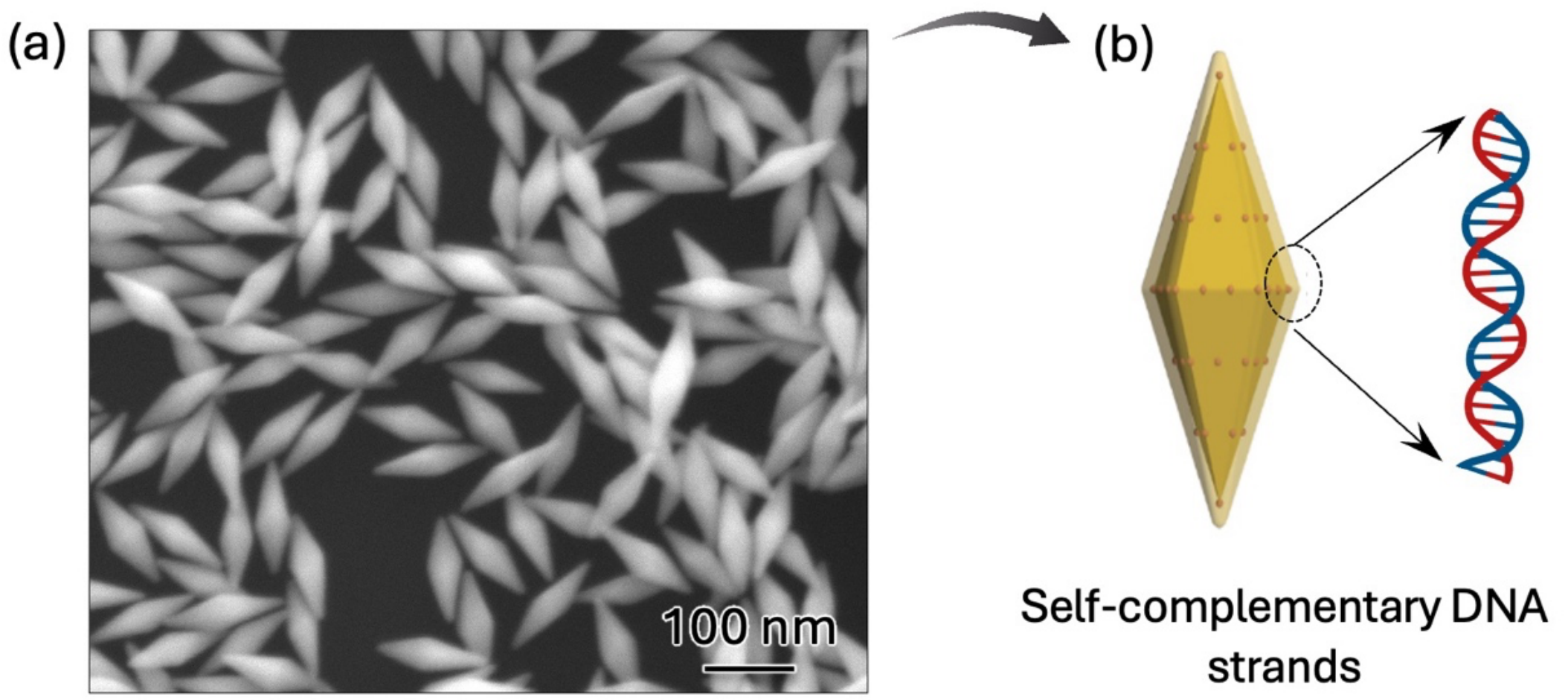

**Figure S1.** (a) SEM image of Au bipyramids of 115 nm. (b) Schematic illustration of the DNA-modified Au bipyramids.

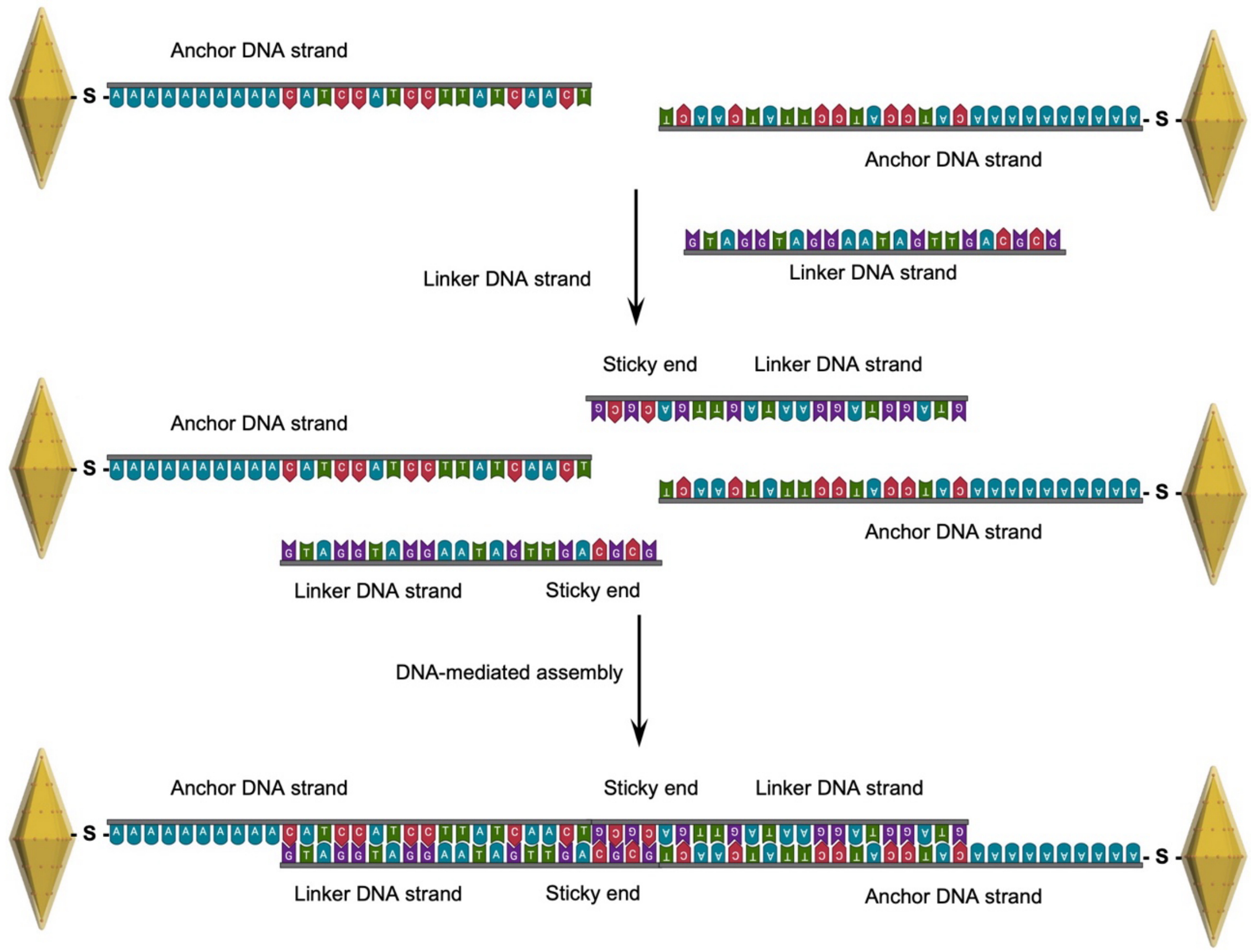

**Figure S2.** Schematic illustration of the DNA-mediated assembly of Au bipyramids using self-complementary strands. Anchor DNA strands are firstly modified on the surfaces of the Au bipyramids though the Au-S bonds. Linker DNA strands are then added, one end of which will bind to the anchor strands while another sticky end (GCGC) are self-complementary.

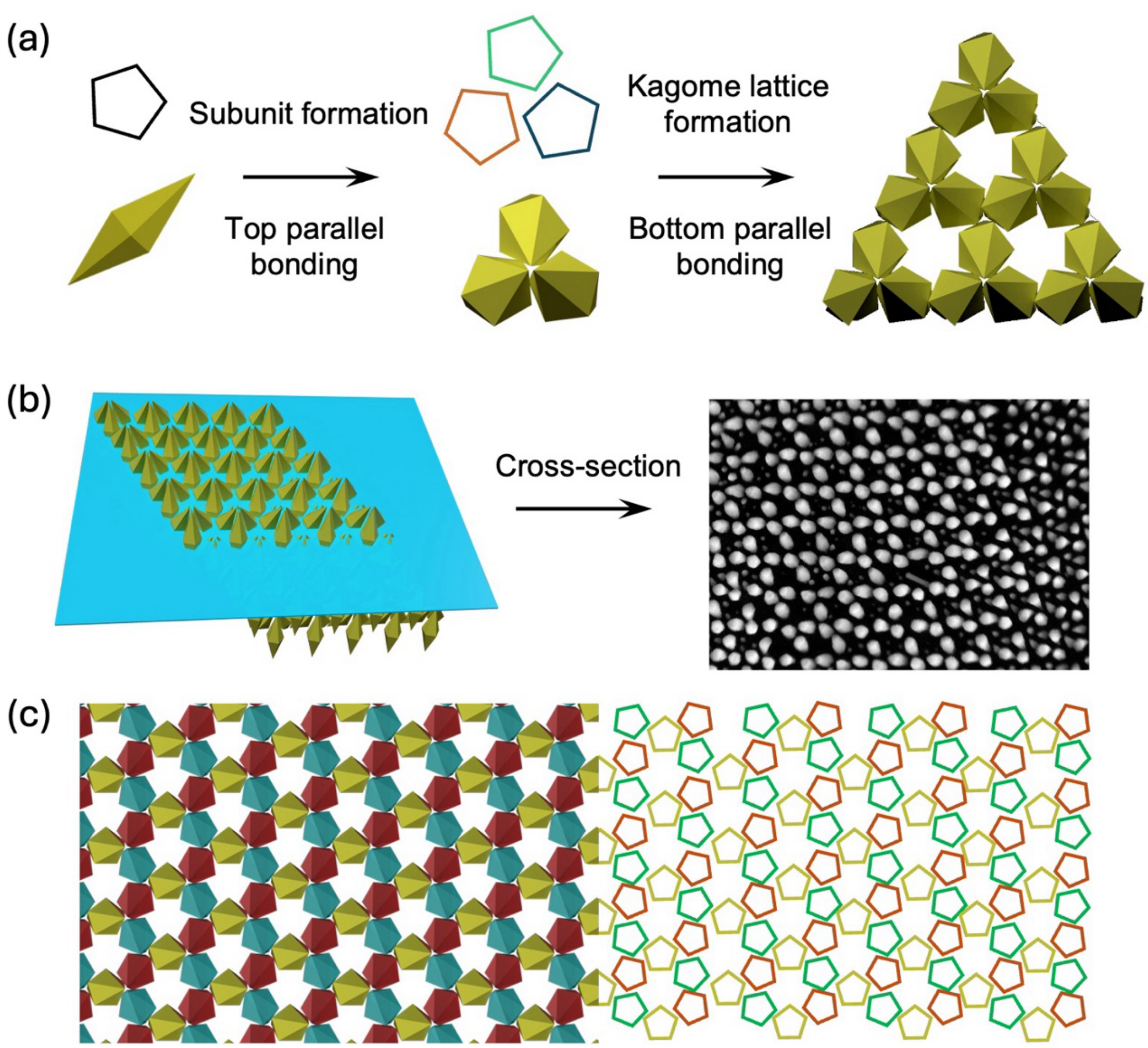

**Figure S3.** Bonding of bipyramids and formation of kagome lattices. (a) Schematic illustration of top and bottom parallel bonding. (b) Schematic illustration of preparing cross section of the kagome lattices and the corresponding TEM images showing the assembly of Au bipyramids into kagome lattices. (c) Schematic illustration of the positional and orientational orders of the Au bipyramids in the kagome lattices.

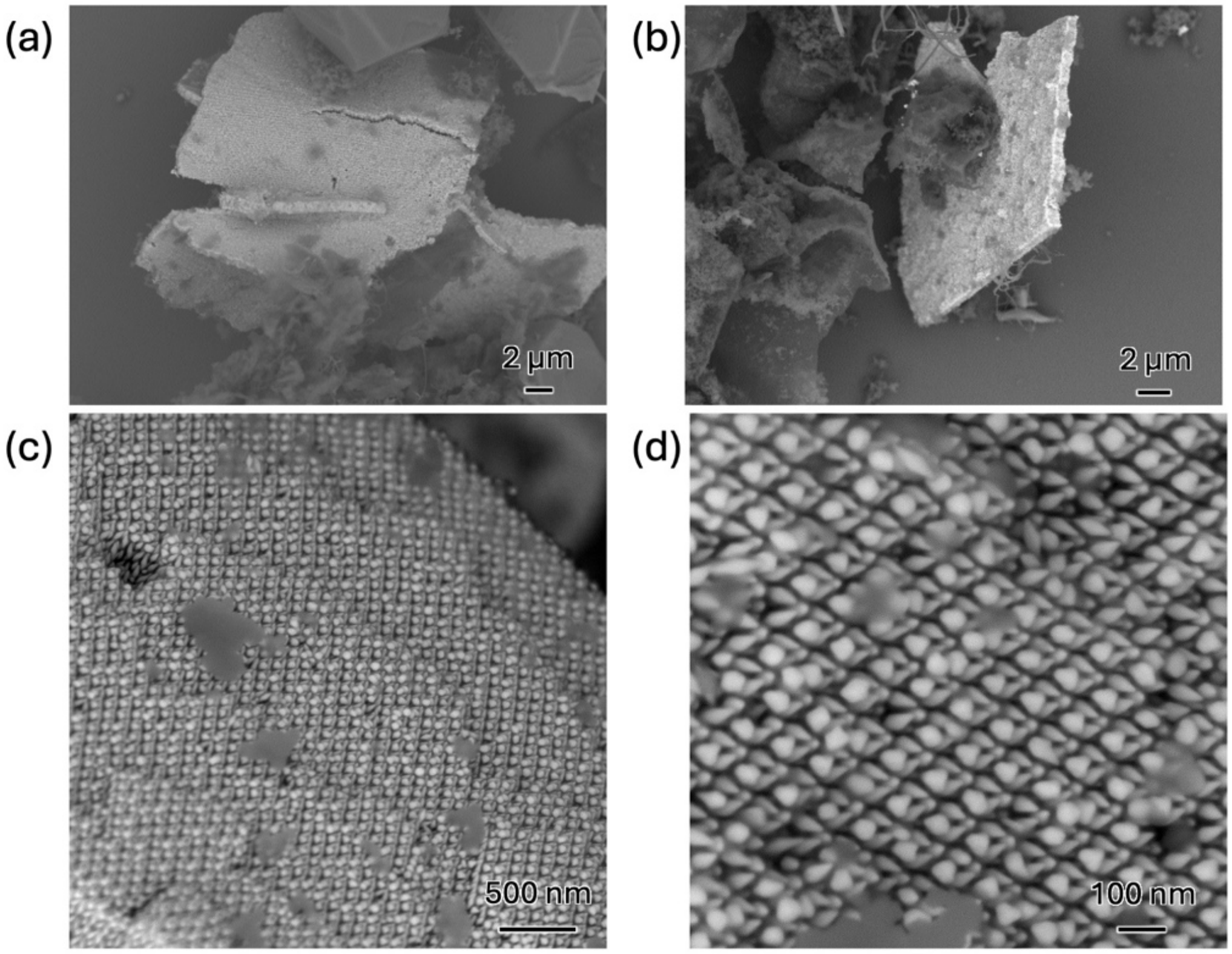

**Figure S4.** (a-b) Low-magnification SEM images of 2D layered colloidal crystals. (c-d) High-magnification SEM images of the 2D layered colloidal crystals.

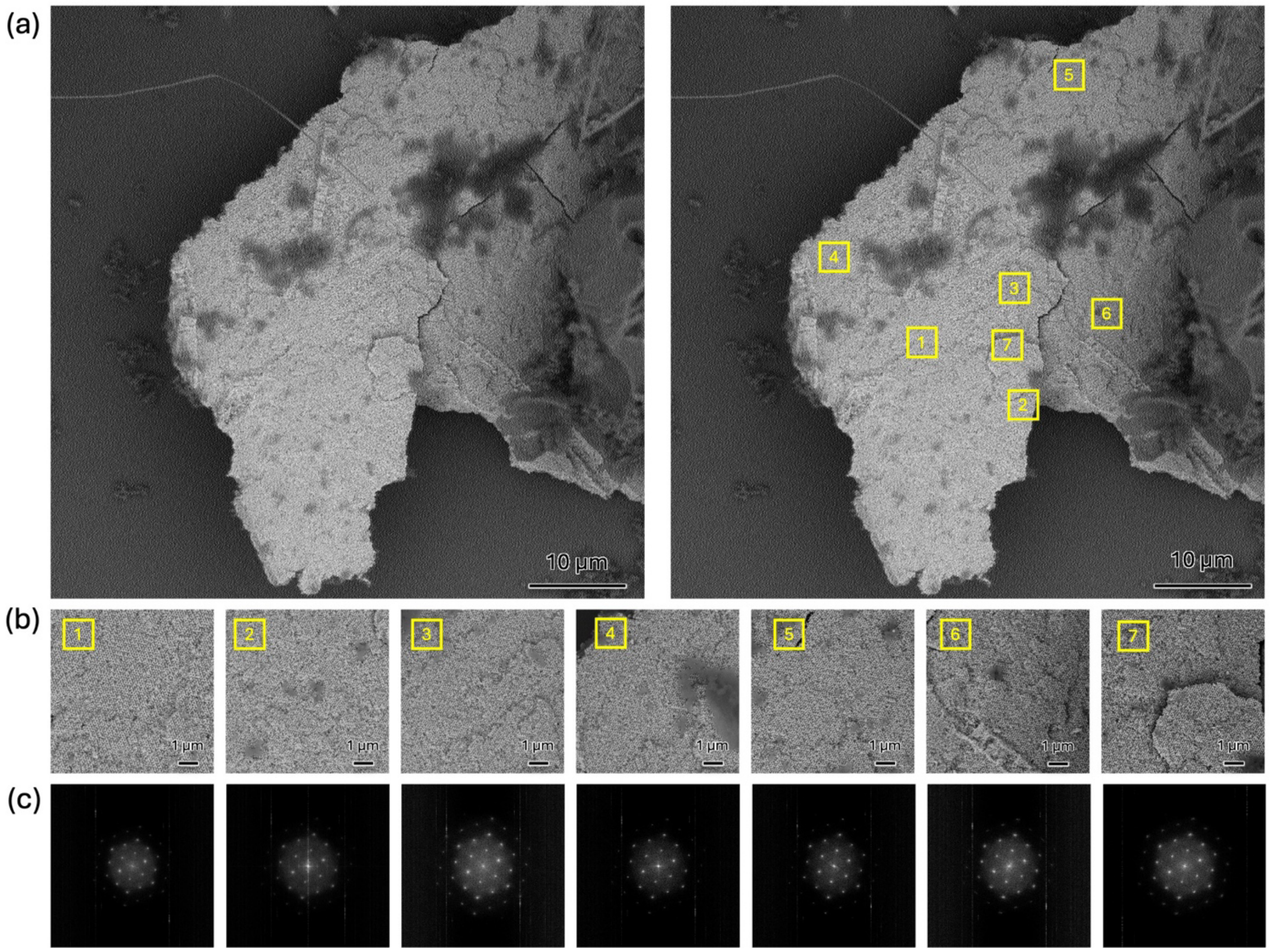

**Figure S5. (a)** SEM images of one colloidal crystal. **(b)** SEM images and **(c)** the corresponding FFT of the SEM images, showing the crystal orientation in different regions. The uniform and consistent diffraction patterns suggest single crystallinity.

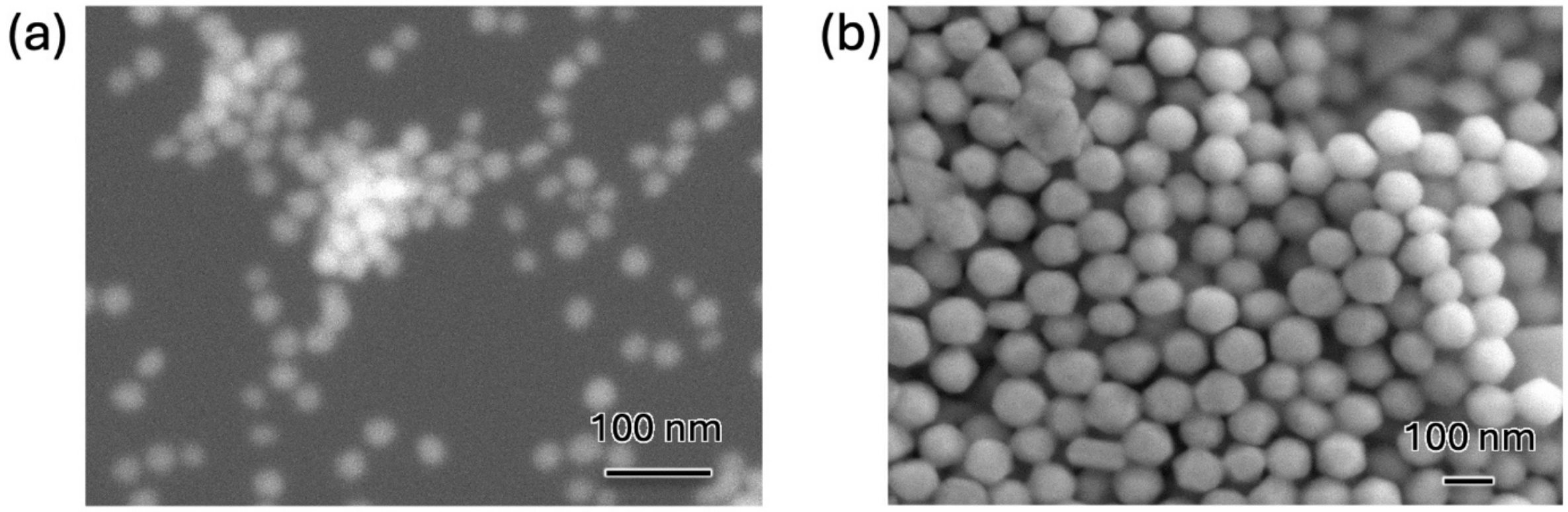

**Figure S6.** SEM images of Au nanospheres of **(a)** 20 nm and **(b)** 80 nm.

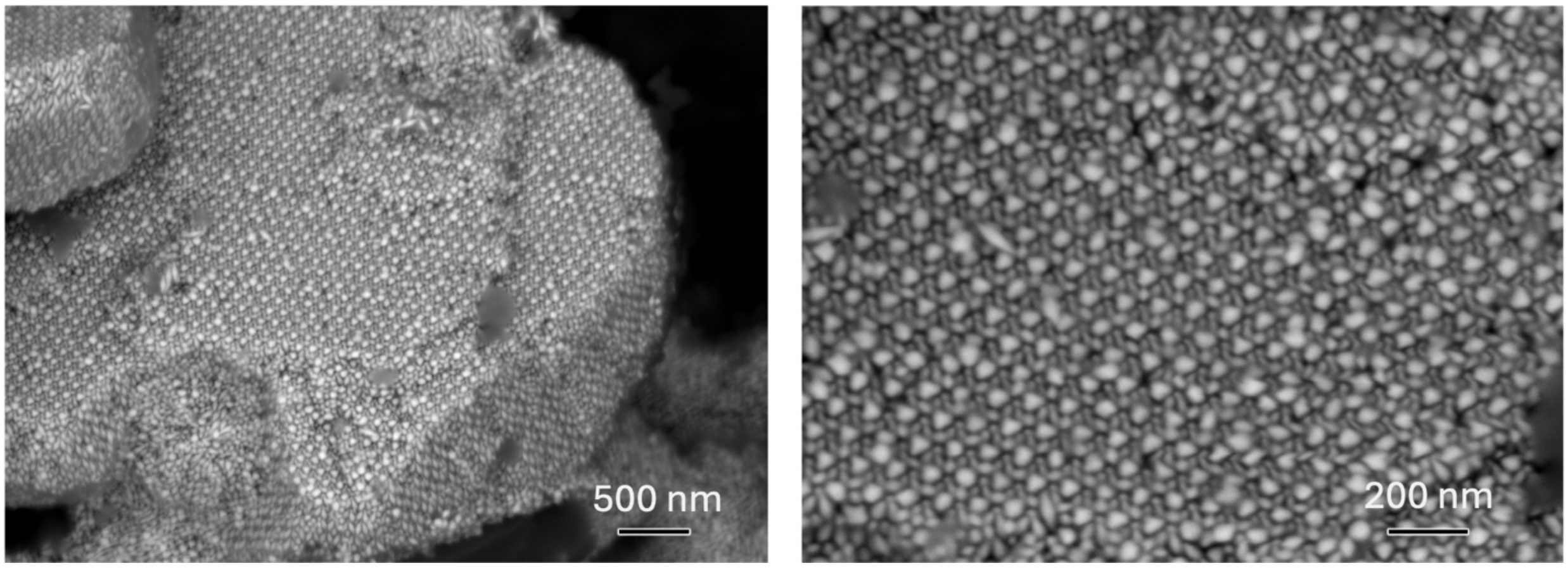

**Figure S7.** SEM images of colloidal crystals by co-assembling 20-nm, 80-nm Au nanospheres with Au bipyramids, demonstrating high selectivity in surface registration of small and large nanospheres.